\documentclass[12pt]{article}

\usepackage{scicite}

\usepackage{times}
\usepackage{hyperref}
\usepackage{graphicx,amsmath,amssymb,xcolor}

\newenvironment{sciabstract}{%
\begin{quote} \bf}
{\end{quote}}

\title{Mining Google and Apple mobility data: \\  ~  \\
Temporal Anatomy for COVID-19 Social Distancing}

\author
{Corentin Cot,$^{1,2}$  Giacomo Cacciapaglia,$^{1,2\ast}$  Francesco Sannino$^{3,4}$\\
\\
\normalsize{$^{1}$Institut de Physique des deux Infinis de Lyon (IP2I),}\\
\normalsize{UMR5822, CNRS/IN2P3, F-69622, Villeurbanne, France}\\
\normalsize{$^{2}$University of Lyon, Universit{\' e} Claude Bernard Lyon 1,  F-69001, Lyon, France}\\
\normalsize{$^{3}$CP3-Origins \& the Danish Institute for Advanced Study, Danish IAS,}\\
\normalsize{University of Southern Denmark, Campusvej 55, DK-5230 Odense M, Denmark}\\
\normalsize{$^{4}$ Dipartimento di Fisica ``E. Pancini'', Universit\`a di Napoli Federico II \& INFN sezione di Napoli,}\\
\normalsize{Complesso Universitario di Monte S. Angelo Edificio 6, via Cintia, 80126 Napoli, Italy}\\
\\
\normalsize{$^\ast$To whom correspondence should be addressed; E-mail:  g.cacciapaglia@ipnl.in2p3.fr}
}

\date{}
\begin{document}

\baselineskip24pt

\maketitle

\begin{sciabstract}
 We employ the Google and Apple mobility data to identify, quantify and classify different degrees of social distancing and characterise their imprint on the first wave of the COVID-19 pandemic in Europe and in the United States. We identify the period of enacted social distancing via  Google and Apple data, independently from the political decisions.  Interestingly we observe a general decrease in the infection rate occurring two to five weeks after the onset of mobility reduction for the European countries and the American states. 
\end{sciabstract}

COVID-19 has disrupted our way of living with long lasting impact on our social behaviour and the world economy. At the same time, differently from earlier pandemics, a very large amount of data has been collected~\cite{wellenius2020impacts,jiaa491,islind2020changes} thanks, also, to our smartphone dominated society.  Smartphones run mobility applications, such as Google and/or Apple Maps, that help humans navigate. The mobility information stemming from these apps has been harvested by Google and Apple, which have subsequently made it publicly available on the following websites: \href{https://www.google.com/covid19/mobility/}{Google} and
 \href{https://www.apple.com/covid19/mobility}{Apple}. 
 
 In this paper we mine these data to quantify and characterise the effects of social distancing measures enacted by various European countries and American states. An early study of mobility effects on the pandemic evolution in China can be found in Ref.~\cite{Lai2020}. The Google mobility data, in Google wordings, {\it show movement trends by region, across different categories of places.} As categories we will use ``Residential'' and ``Workplace'', which best describe the change in people's behaviour after the implementation of social distancing measures with respect to a baseline day. The latter is defined, according to Google, as the median value from the 5-week period from the 3rd of January to the 6th of February, 2020, predating the wide spread of the virus in Europe. The data show how visitors to (or time spent in) categorised places changed with respect to the baseline day.  For Apple, the available mobility data represent a relative volume of direction requests per country/region, sub-region or city, compared to a baseline volume defined on the 13th of January, 2020. We will be using, from Apple, information about ``Driving'' and ``Walking'', assuming they represent the time spent by people away from home.  For the United States (US), only ``Driving'' data are available.
  
 Another set of data relevant for this work is related to the virus spreading dynamics, which we take from the website \href{https://ourworldindata.org/}{https://ourworldindata.org/}. We normalise the data of each country as cases per million inhabitants.

  The data relative to the total number of infected cases are effectively parameterised using the High Energy Physics inspired formalism introduced in  \cite{DellaMorte:2020wlc}, dubbed \emph{epidemic Renormalisation Group} (eRG).  The approach has been generalised to take into account the spreading dynamics across different regions of the world in \cite{Cacciapaglia:2020mjf} and the evolution of the second wave pandemic across Europe in \cite{cacciapaglia2020second}. The advantage of the eRG formalism resides in the limited number of coefficients needed to classify the spreading dynamics for each country. 
  More complicated models have been used in the literature to study the effect of non-pharmaceutical interventions, including mobility, for Europe~\cite{Flaxman2020} and the US~\cite{kabiri2020different,nielsen2020clustering,fellows2020covid19,hong2020exposure,vanni2020epidemic}, with the latter mostly focusing on local communities.
 
  Without further ado, following   \cite{DellaMorte:2020wlc,Cacciapaglia:2020mjf}, we introduce $\alpha(t)$ below
\begin{equation}
    \alpha(t) = \text{ln}\left(\mathcal{I}(t)\right) \ ,
\end{equation}
where $\mathcal{I}(t)$ is the total number of infected cases per million inhabitants in a given region and $\ln$ indicates its natural logarithm.  The function $\alpha(t)$ turns out to be well described by the following logistic function:
\begin{equation}
    \alpha(t) = \frac{ae^{\gamma t}}{b + e^{\gamma t}} \ .
\label{eq:logistic}
\end{equation}
Here, $a$ represents the logarithm of the final number of infected cases per million inhabitants, $b$ denotes the temporal shift from the start of the pandemic and $\gamma$ measures the  \emph{flatness} of the curve of the number of new infected cases. Here, and in the following, we will measure the time $t$ in weeks, so that $\gamma$ is measured in inverse weeks. 
It was argued in \cite{DellaMorte:2020wlc,Cacciapaglia:2020mjf} that, aside from the trivial temporal shift provided by $b$ and for the first wave of the pandemics, two numbers are sufficient to characterise the evolution of the number of infected cases per each region, i.e. $a$ and $\gamma$.  This fact  helps studying the correlation between mobility data and the virus spreading dynamics for each region. By going beyond the previous parameterisation,  we will discover a finer temporal structure directly related to the effects of the imposed lockdown and social distancing measures in the different regions. 

In this work we focus on a selection of European countries and all of the US states. In Europe, we considered countries with more than 3 million inhabitants and for which the data were available. 
Note that we will only consider the period during which the first wave of the COVID-19 was raging in Europe and in the US.

Using Google and Apple data, we provide a rationale to identify the timing of the social distancing measure actualisation in each region. European countries and US states adopted different degrees of social distancing measures during the first wave of the COVID-19 pandemic. Moreover the severity of the measures changed during the spreading of the epidemic within each region of the world.  This is why we defined the beginning of the impact of social distancing measures in terms of the reduction in the mobility of individuals, rather than on political decisions.

We mine Google's Residential and Workplace mobility data since they show movement trends across different places compared to a reference period before the implementation of any measure. The Residential and Workplace data are best suited to quantify when and to what extent people reduced their mobility and increased social isolation.  
Similarly, for Apple, we choose Driving and Walking data for Europe and Driving for the US states, expressing them in terms of a percentage reduction. Note that the Apple data refer to variations in the number of searches done on the Maps app, more details to be found on the website  \href{https://www.apple.com/covid19/mobility}{Apple}.

  \begin{figure}[tb!]
\includegraphics[scale=.45]{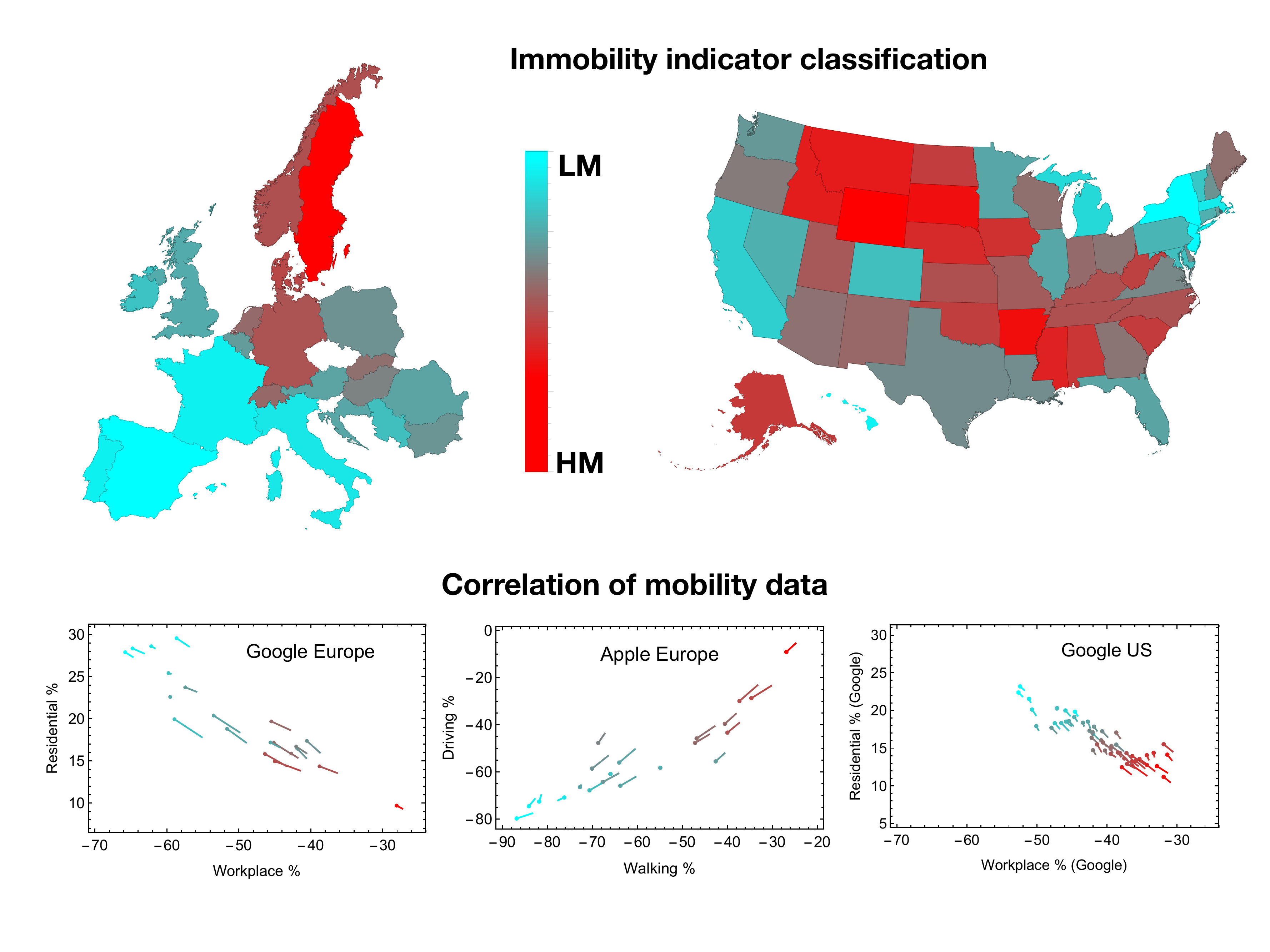}
\caption{{\bf The COVID-19 Mobility Map for Europe and the US}. The two maps represent respectively the European and US states with different shades of mobility from the highest (HM) in bright red to the lowest (LM) in cyan. At the bottom of the figure there are three  tadpole-like plots showing correlations between the four mobility reduction categories: Residential and Workplace from Google, Driving and Walking from Apple. The head of the tadpoles correspond to the average over 6 weeks after social distancing begins, while the tail indicates a 8 week average. The colour code in the three plots reflects the maps one. The maps are drawn with Wolfram Mathematica.}
\label{Res-Work}
\end{figure}

We define an immobility indicator $\diagdown{\hspace{-0.4cm} M}$, as described in the section Methods in the supplementary material, in terms of an average mobility reduction in the chosen categories.  The average is taken over six weeks after the beginning of social distancing. This indicator allows to sort the European countries and the American states based on the hardness of social distancing. 
We also define regions with the highest rate of mobility reduction (\emph{low mobility}, LM) and regions with the least reduction (\emph{high  mobility}, HM), for Europe and the US separately.  The results are shown in Fig.~\ref{Res-Work}, were we indicate the LM regions in cyan and the HM regions in red, with a colour gradient representing different shades of mobility being proportional to the value of the indicator. For Europe, the countries with the smallest mobility grossly correspond to those that imposed a lockdown, while the highest mobility country is Sweden, where no measures were imposed. Nevertheless, even for Sweden the mobility data show a significant variation that allows us to define the beginning of social distancing despite the political decisions. Similarly for the US states, the  lowest LM corresponds to states in the North-East, California and Hawaii, which imposed lockdown measures. We also noticed that the beginning of the measures, as defined by the mobility data in the US, corresponds to the dates when the schools were closed in each state~\cite{wellenius2020impacts}.

To validate our conclusions, at the bottom of Fig.~\ref{Res-Work} we show the correlations between Google and Apple mobility data for Europe (left and central plot) and the US (right plot). Each region is represented by a tadpole-like symbol, with the head corresponding to the 6-week average and the tail to the 8-week average. We label each country and state by using the same colour code as in the maps. The plots show a clear correlation between the percentage change in each category. We also checked that the same correlation persists when comparing Google to Apple categories.

We now analyse possible correlations between mobility data and the parameters of the logistic function $\alpha(t)$ such as the infection rate $\gamma$ and the log of the total number of infected cases $a$. To our surprise we find that $\gamma$ is uncorrelated to the degree of mobility reduction. This implies that mobility changes have little impact on the velocity of diffusion of the disease.  Of course, mobility data only capture one aspect of the social distancing, thus they do not offer a complete picture of the situation in various regions. 
This surprising finding can be interpreted in various ways. On the one hand, the result may imply that the main factor behind a reduction of $\gamma$ could lie in the behaviour of individuals in social occasions (mask wearing, proximity, greeting habits, to mention a few); on the other hand, it is quite possible that the value of $\gamma$ does not represent the effect of the social distancing measures, as it derives from a global fit over a wide timescale. In other words, the fit values include both the measure and the pre-measure periods.

\begin{figure}[tb!]
\includegraphics[scale=0.45]{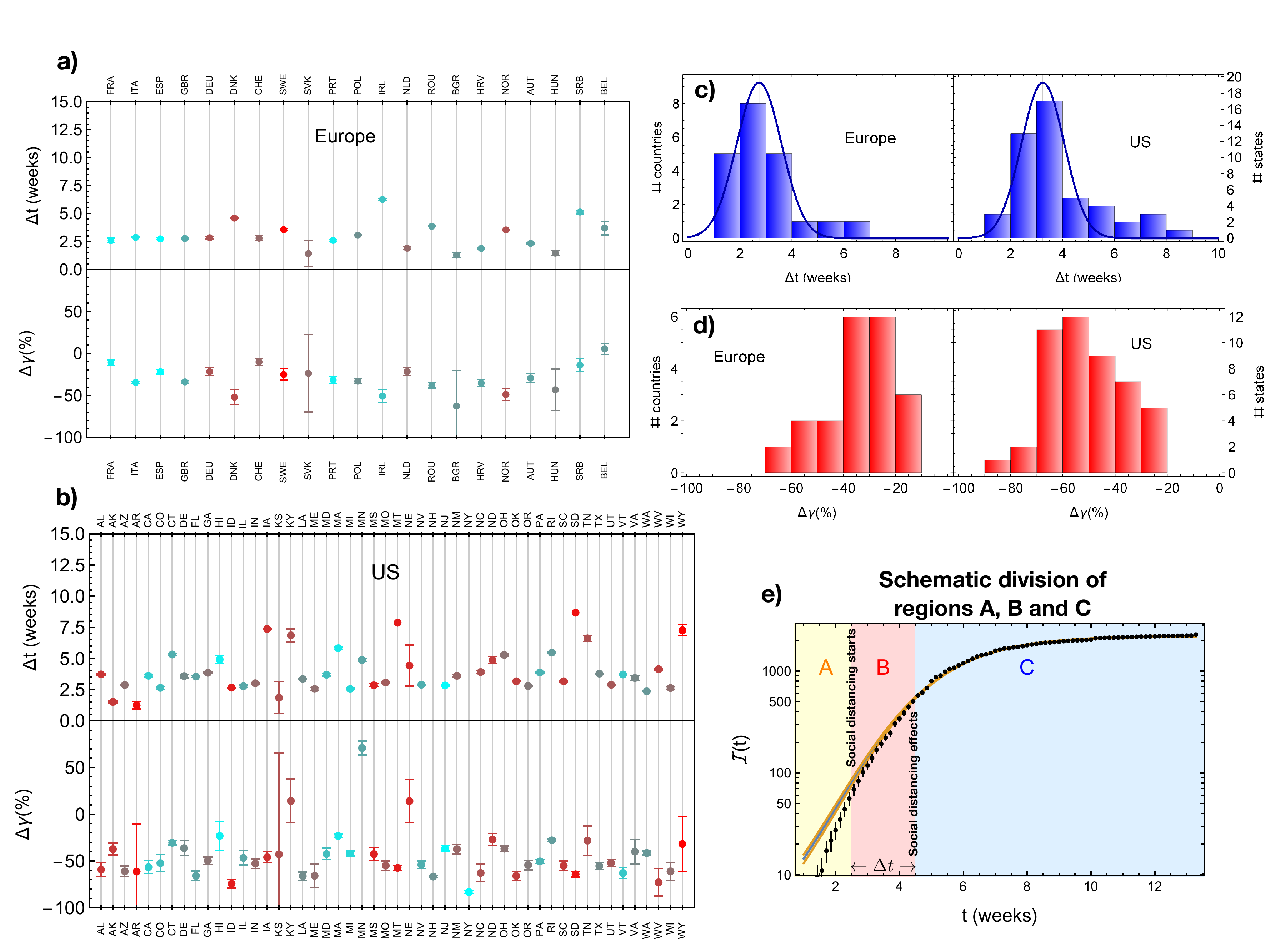}
\caption{{\bf Temporal anatomy of COVID-19 social distancing effects}. In panel a), we show $\Delta t$ and the percentage variation $\Delta \gamma$ in the infection rate for the European countries considered in this study. In panel b), we show the same for all the US states. In panels c) and d) we display the same results in the form of histograms, for Europe and the US separately, highlighting that $\Delta t$ clusters around similar values. In panel e), we illustrate the subdivision of the first wave epidemic curve in three temporal regions: A before social distancing as defined via mobility data occurs, B until an effect is observed in the epidemic curve as a change in $\gamma$, and C covering the later times. $\Delta t$ equals to the duration of the period B. }
\label{fig2}
\end{figure}

To push further the analysis, we now explore whether social distancing measures (as defined via the Apple/Google mobility data) lead to distinct temporal patterns in the European countries under study and the American states. In the eRG approach, $\gamma$ is the natural parameter to use for this task. We assume that, after the measures are enacted, there are two distinct temporal regions describing the time dependence of the number of infected cases. These two regions, B and C in the illustrative plot in Fig.~\ref{fig2}e, are naturally described by two different gammas. We confirm that such an analysis is possible via a MonteCarlo analysis. 
We then move to the actual data and discover that two distinct temporal regions with their own gammas do emerge for several regions. In Fig.~\ref{fig2} we show the outcome of the fit to the data in terms of the time interval $\Delta t$ between the beginning of social distancing  and its effects measured when the infection rate $\gamma$ changes.  We discover that most countries display a similar $\Delta t$. By fitting the distributions in Fig.~\ref{fig2}c to a gaussian, we find that to the two sigma level we have $\Delta t = 2.7 \pm 1.7$ weeks for Europe and $\Delta t = 3.3 \pm 1.6$ weeks for the US. The high compatibility of the two ranges shows the emergence of a universal time scale for  social distancing to be effective. 

Another important result is the general and strong reduction of the infection rate measured within and after $\Delta t$ both for Europe and the US, as shown in the left panels of Fig.~\ref{fig2} and summarised by the red histograms of Fig.~\ref{fig2}d.

\section*{Discussion}

We analysed the mobility data released by Google and Apple to quantify the effects of social distancing on the COVID-19 spreading dynamics in Europe and in the US. We were able to classify different shades of social distancing measures for the first wave of the pandemic. After identifying the countries according to their level of immobility, we  observed a strong decrease in the infection rate occurring two to five weeks after the onset of mobility reduction for the countries studied here. Another important result is the discovery of a universal time scale after which social distancing starts showing its impact. 

We have also provided an actual measure of the impact of social distancing for each region,  showing that the effect amounts to a reduction by $20$\% $-$ $40$\% in the infection rate for most countries in Europe and $30$\% $-$ $70$\% in the US. This is, to the best of our knowledge, a first global and direct measure of the impact of social distancing. Interestingly, even  countries that did not impose political measures, like Sweden, show a reduction of the infection rate similar to the ones experiencing a lockdown, suggesting that a certain degree of social restrain occurred regardless of the political decisions.  
Our results are compatible with early analysis of local social distancing measures taken in China~\cite{Lai2020}, where mobility data inter-cities from \href{https://qianxi.baidu.com/}{Baidu} was used within a compartmental model.

Using smartphone based open-source mobility data, we showed that it is possible to provide a temporal anatomy of social distancing.  We discovered the emergence of a characteristic time scale related to when social distancing effects have a measurable impact. This timing can also be used to quantify the impact of social distancing by determining the variation in infection rate per country. Finding similar reduction, however, does not imply that the countries have a similar number of infected cases per million inhabitants. It simply means that there has been a change in social behaviour. The result of this study, based on the simple eRG approach, lays the basis for an effective tool for  the authorities to evaluate the timing and impact of the imposition of social distancing measures, in particular related to movement restrictions.

\bibliographystyle{Science}

\bibliography{biblio}

\section*{Acknowledgements}

G.C. and C.C. acknowledge partial support from the Labex-LIO (Lyon Institute of Origins) under grant ANR-10-LABX-66 (Agence Nationale pour la Recherche), and FRAMA (FR3127, F\'ed\'eration de Recherche ``Andr\'e Marie Amp\`ere'').

\section*{Author contribution}

This work has been designed and performed conjointly and equally by the authors.
G.C., C.C. and F.S. have equally contributed to the writing of the article.

\section*{Competing interests}

The authors declare no competing interests.

 \section*{Materials and Methods}

\paragraph*{Immobility indicator.}

European countries and US states adopted different degrees of social distancing measures during the first wave of the COVID-19 pandemic. Moreover the severity of the measures changed during the spreading of the epidemic within each country or state. Rather than classifying the countries based on their political choices, we  use the mobility data provided by Google and Apple as indicators of the effective hardness of the measures. 

 \begin{figure}[tb!]
\includegraphics[width=7cm]{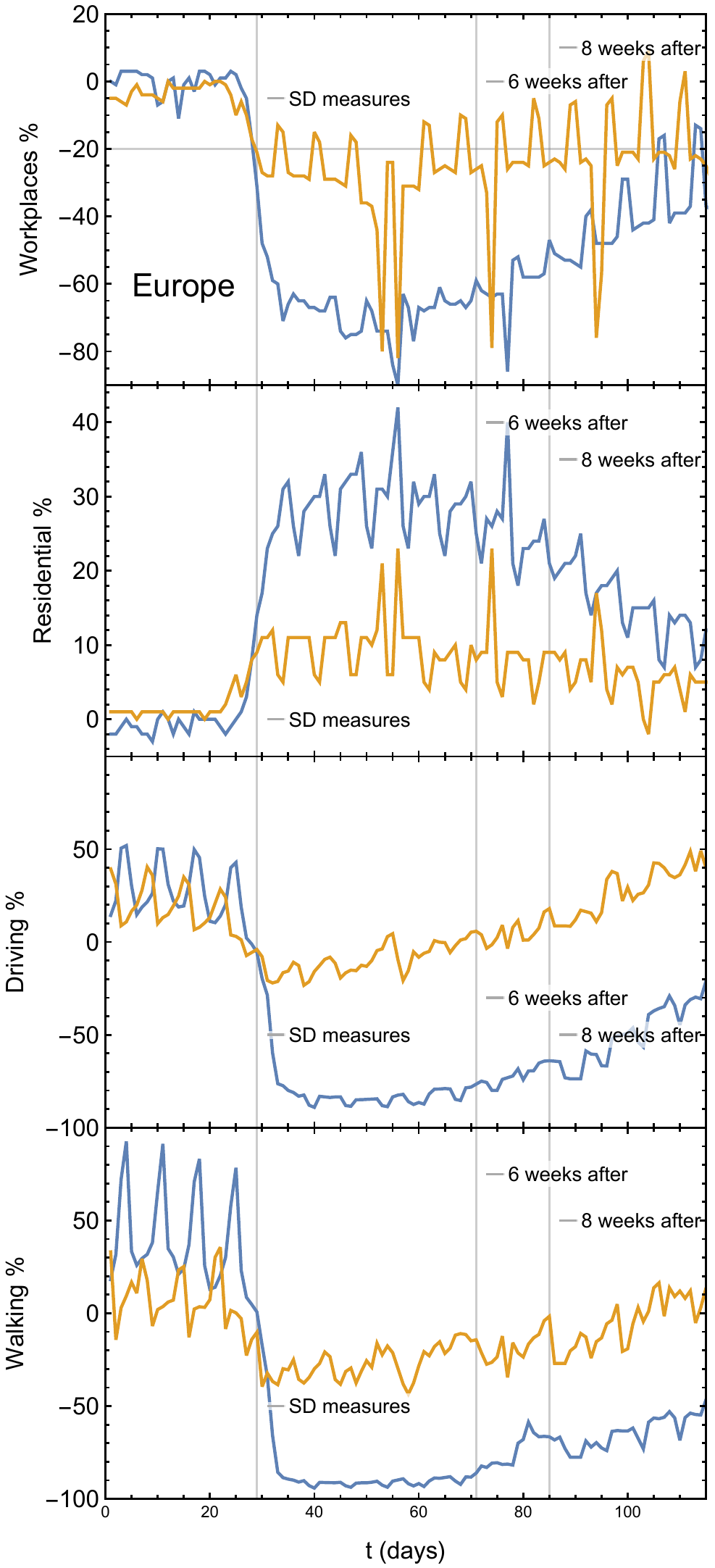}  \hspace{1cm}
\includegraphics[width=7cm]{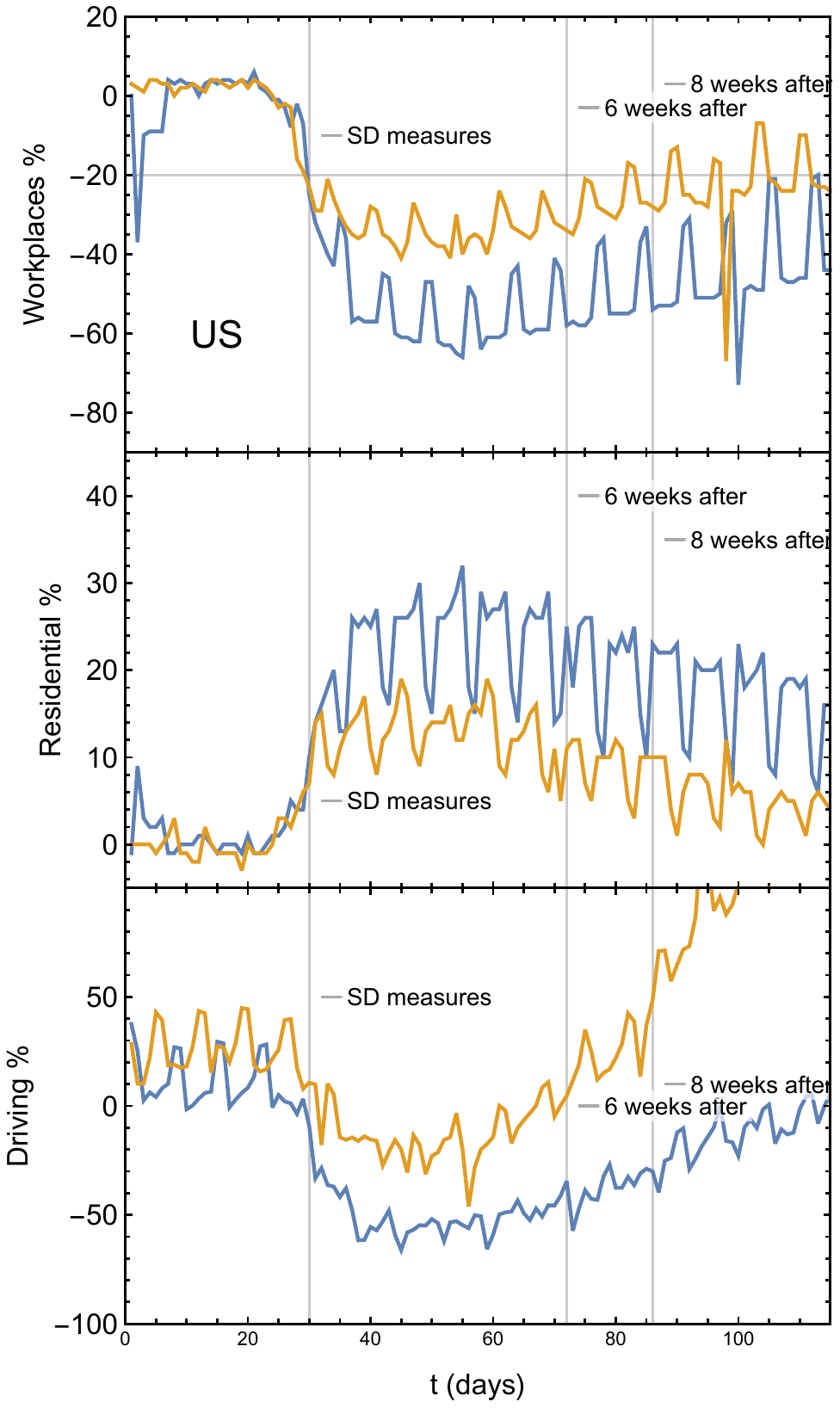}
\caption{{\bf Raw Google and Mobility data.} In these plots we show a sample of raw Google and Apple mobility data used in this work, for Europe (left) and the US (right). The time scale is shifted so that the beginning of the social distancing, defined by a $20\%$ drop in Google's Workplace and indicated by the first vertical grey line, coincides for all countries and states. The other two vertical lines mark the end of the 6 and 8 week averaging periods respectively. We show the respective HM region in orange and the LM one in blue: for Europe, Sweden (orange) and Spain (blue); for the US, Wyoming (orange) and New York (blue).}
\label{Fig:Raw}
\end{figure}

To find a measure for the \emph{immobility} of a given population during the social distancing period, we define an average percentage variation for each of the four categories: Residential and Workplace for Google and Driving and Walking for Apple (only Driving is available for US states). 
For both mobility datasets, the percentage variations are defined with respect to a reference date or period predating the exponential growth of the infection cases. The data are typically very jugged, as illustrated in Fig.~\ref{Fig:Raw}, mainly due to strong variations over the weekend. Furthermore, the mobility data feature a sharp decrease followed by a slow return to the pre-COVID-19 average. Taking into account this behaviour, it is necessary to define an average over several weeks, which would allow us to associate a single number to each category and region.

Firstly, one needs to properly define the beginning of the social distancing period for each region: we choose to identify it with the time when Google Workplace percentage first drops by 20\% (at this time, typically, all mobility indicators have shown a significant variation). The ending of the measure period is harder to identify, as the social distancing measures have always been lifted progressively ({\it 3\/}): this appears in the mobility data, as the curves gradually return to zero, i.e. to the reference period levels, or even above.  Thus, we decided to fix the same averaging period  for all the regions we considered. To test the robustness of our conclusions, we determine the outcome for two choices: 6 and 8 weeks after the effective beginning of the measures. The tadpole-like plot at the bottom of Fig.1 in the main text demonstrate that the duration of the averaging period, while changing the value of the mobility reduction, does preserve the overall trend. In the following, therefore, we will use the 6-week average as our benchmark.

\begin{figure}[tb!]
\includegraphics[scale=0.8]{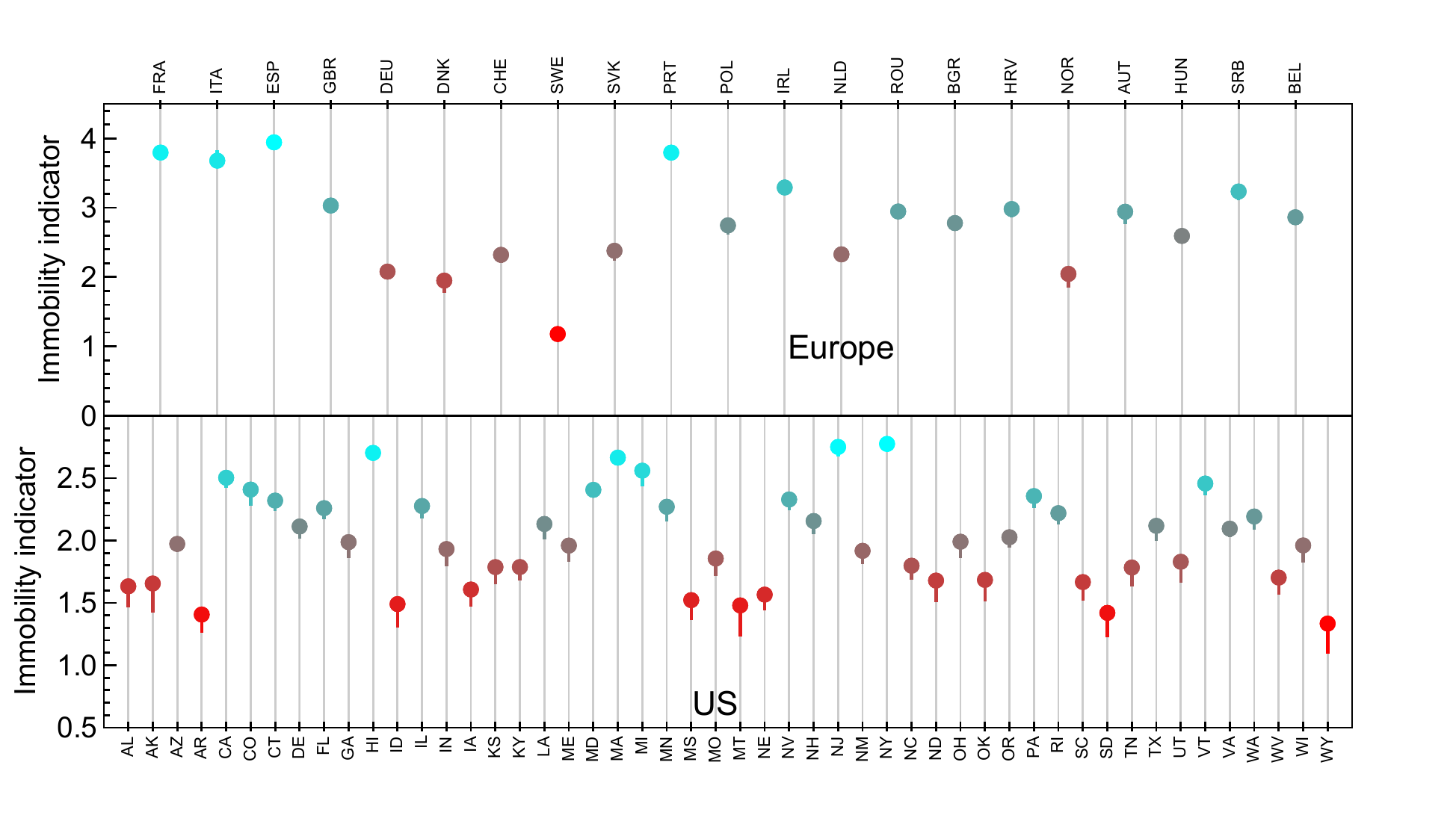} 
\caption{{\bf Immobility indicator for the European countries and the US states.} Values of the immobility indicator $\diagdown{\hspace{-0.4cm} M}$ for Europe (top) and the US (bottom). The colour code corresponds to the ranking of each European country and each US state,  matching the one used in Fig.1 of the main text. }
\label{Fig:Imm}
\end{figure}

To be able to classify the countries based on their immobility, we further define an immobility indicator as
\begin{equation}
\diagdown{\hspace{-0.4cm} M} (\mbox{region}) = \sum_{j=\mbox{cat.}} \;  \frac{|p_j (\mbox{region})|}{\mbox{max} [ |p_j| ]}\,,
\end{equation} 
where $|p_j (\mbox{region})|$ is the absolute value of the percentage variation in each category (labelled by $j$). For each category, we divide by the maximal value observed in the pool. Note that for European countries we have 4 categories, so that $\diagdown{\hspace{-0.4cm} M} < 4$, while for the US states we have 3 categories, so that $\diagdown{\hspace{-0.4cm} M} < 3$.  
We use this indicator to rank the European countries and the American states from the ones with \emph{high mobility} (HM) --  small $\diagdown{\hspace{-0.4cm} M}$ -- to the one with \emph{low mobility} (LM) --  large $\diagdown{\hspace{-0.4cm} M}$.
The values of the immobility indicator we obtain for the European countries under study and US states are shown in Fig.~\ref{Fig:Imm}. The colour code ranges from the highest mobility region in bright red to the lowest mobility one in cyan, with gradient proportional to the value of the immobility indicator.

\paragraph*{Comparing the virus spreading parameters with mobility data.}

The epidemic evolution of the first wave of the CODIV-19 pandemic can be effectively characterised by two parameters: the infection rate $\gamma$ and the logarithm of the final number of total infected cases $a$ ({\it 5 \/}), measured per million inhabitants. We remark, however, that it is risky to compare the number of infected for different regions due to the different procedures used when identifying the positive cases, and the different testing rates and strategies. Thus, we assign more physical meaning to the infection rates $\gamma$, which give an accurate temporal characterisation of the epidemic diffusion in each region.

\begin{figure}[tb!]
\includegraphics[scale=0.7]{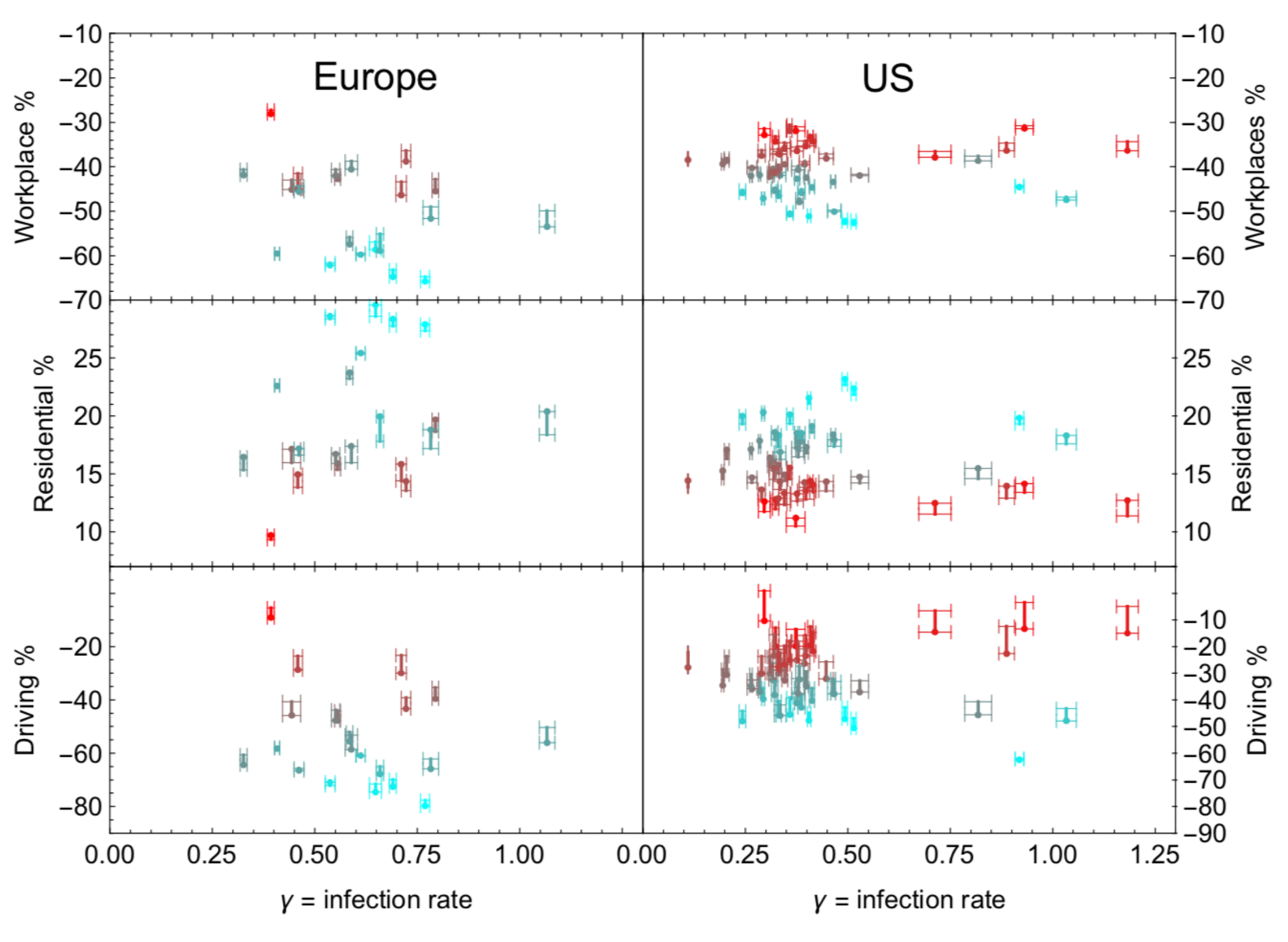} 
\caption{{\bf Infection rate compared to the mobility data}. Racecar plots showing the fitted infection rates $\gamma$ versus the Google/Apple mobility categories. The vertical segment indicates the difference between 6 week (dot) and 8 week averages; the horizontal bars indicate the fit error on $\gamma$.}
\label{Fig:Goo}
\end{figure}

It is, therefore, natural to hypothesise that regions with higher mobility may have a faster diffusion rate of the infection, i.e. larger values of $\gamma$.  
To test this hypothesis, in Fig.~\ref{Fig:Goo}, we show the Workplace, Residential and Driving reductions versus the infection rates for the European countries in this study and the US states. To each country or state is associated a racecar-like symbol: the pilot seat (dot) corresponds to the 6-week average, while the tail to the 8-week average. Furthermore, the side bars indicate the error from the fits of the epidemic data. The colour codes match the immobility indicator defined above. The data used to generate the plots in Fig.~\ref{Fig:Goo} are reported in Tables~\ref{T1} and \ref{T3}, where we only report the mobility averages over 6 weeks.

Surprisingly, the data do not reveal any particular correlation between the values of $\gamma$ and the mobility data. As explained in the main text, this result can be interpreted in various ways. One possibility, which we will test in the following section, is that the $\gamma$ from the fit of the first wave is not the most appropriate measure, as it averages over the infection diffusion before and after the mobility reduction occurs.

\paragraph*{Testing the two-gamma hypothesis.}

 \begin{figure*}[tb!]
 \includegraphics[scale=.9]{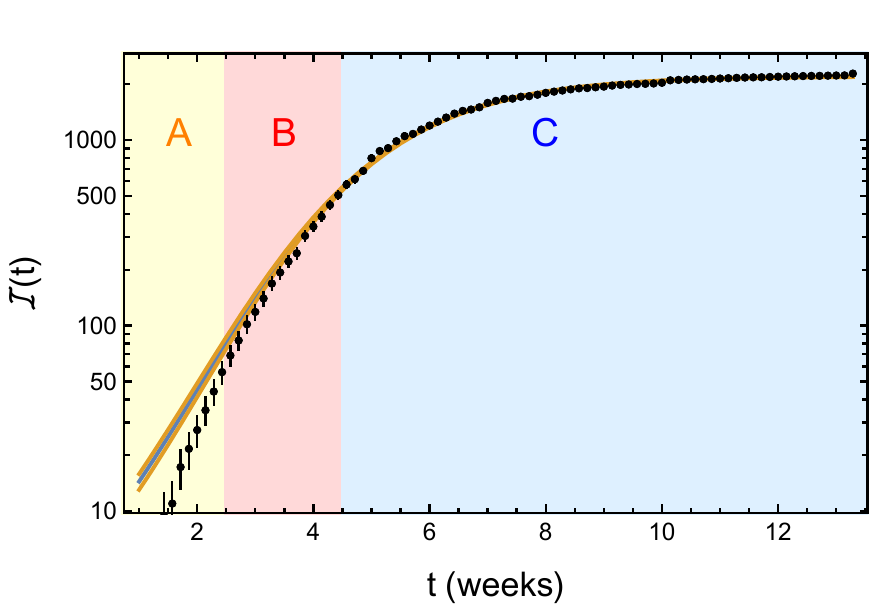} \hfill \includegraphics[scale=.75]{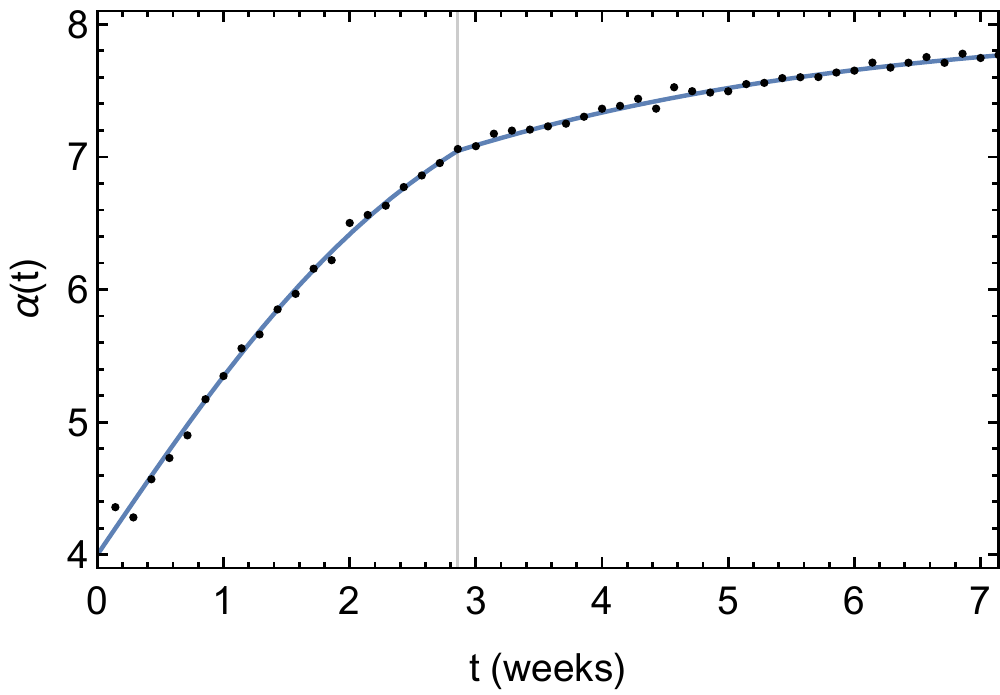}
\caption{{\bf Temporal anatomy of the first wave epidemic data}. Left panel: schema of the 3 temporal regions defined in the text. A refers to the pre-measure time, B occurs between the start of the social distancing and the change in $\gamma$, C covers the later times, after the measure effects occur. The duration of B is defined as $\Delta t$. Right panel: generating function (solid) and sample of the simulated points for the two-gamma model.}
\label{DataNormal}
\end{figure*}

We subdivide the period of the virus diffusion in 3 parts, as illustrated in the left panel of Fig.~\ref{DataNormal}. Region A extends up to the time when the social distancing starts, $t=0$, as defined from the mobility data; at this point Region B begins extending for a duration $\Delta t$; finally Region C starts at $t=\Delta t$. As the beginning of Region B is determined by the Google/Apple mobility data, we can probe the existence of a change in $\gamma$ by fitting the data in Region B+C with the following function:  
\begin{equation}
    \alpha_{2\gamma}(t) = \left\{
    \begin{array}{ll}
        a\frac{\text{exp}\left(\gamma_B t\right)}{b+\text{exp}\left(\gamma_B t\right)} & \mbox{for } t < \Delta t   \\
        \\
        a\frac{\text{exp}\left(\gamma_C t\right)}{b\,\text{exp}\left((\gamma_C - \gamma_B)\Delta t\right)+\text{exp}\left(\gamma_C t\right)} & \mbox{for } t > \Delta t
    \end{array}
\right.
\label{alpha2}
\end{equation}
that depends on 5 parameters: $a$, $b$, $\gamma_B$, $\gamma_C$ and $\Delta t$. We then extract the values of the 5 parameters by fitting to the data.

We first test the effectiveness of our method by generating a mock set of data based on the function in Eq.~\eqref{alpha2}, where we fix $\gamma_B = 0.7$, $\gamma_C = 0.35$ and $\Delta t = 20$ days. An example of the generated data, overlaid to the generating function, is shown in the right panel of Fig.~\ref{DataNormal}: the points are randomly generated within a one standard deviation region, i.e. $[N_i-\sqrt{N_i}, N_i + \sqrt{N_i}]$, where $N_i$ is the number of cases per day as predicted by the generating function.  We generated 100 independent sets of mock data and fitted them to Eq.~\eqref{alpha2}.  We found that we can determine the value of $\Delta t$ within a range of two weeks. Furthermore, we define the percentage variation of the infection rate as
\begin{equation}
\Delta \gamma = \frac{\gamma_C - \gamma_B}{\gamma_C}\,.
\end{equation}

Having acquired confidence in the method, we now apply it to the real data.  The results of the fits are reported in Tables~\ref{T2} and \ref{T4}.

 \begin{table*}[tb!]
\begin{center}
{\footnotesize \begin{tabular}{ || l |c|c|c|c|c|c|c|c||}
\hline
\multicolumn{9}{|c|}{Fit parameters: first wave and 6-week average mobility variations in Europe}\\
\hline
Country & ISO & $a$ & $\gamma$ & $b$ & Work. & Res. & Driv. & Walk.\\
\hline
France & FRA & $7.641(5)$ & $0.690(8)$  & $1.20(3)\times 10^3$  & $-65$ & $28$ & $-73$ & $-82$\\

Italy & ITA & $8.185(11)$ & $0.536(12)$ & $106(3)$ & $-62$ & $29$ & $-71$ & $-76$\\

Spain & ESP & $8.432(6)$ & $0.769(11)$ & $1.88(5)\times 10^3$ & $-66$ & $28$ & $-80$ & $-87$\\

United Kingdom & GBR & $8.293(8)$ & $0.407(6)$ & $65(2)$ & $-59$ & $23$ & $-58$ & $-55$\\

Germany & DEU & $7.580(7)$ & $0.722(12)$ & $1.58(4)\times 10^3$ & $-39$ & $14$ & $-43$ & $-40$\\

Denmark & DNK & $7.591(13)$ & $0.458(11)$ & $42(2)$ & $-45$ & $15$ & $-29$ & $-35$\\

Switzerland & CHE & $8.142(3)$ & $0.794(8)$ & $2.45(4)\times 10^3$ & $-45$ & $20$ & $-40$ & $-40$\\

Sweden & SWE & $8.07(3)$ & $0.392(9)$ & $52.6(7)$ & $-28$ & $10$ & $-9.1$ & $-27$\\

Slovakia & SVK & $6.00(6)$ & $0.44(2)$ & $139(7)$ & $-45$ & $17$ & $-46$ & $-47$\\

Portugal & PRT & $7.90(1)$ & $0.648(14)$ & $9.5(4)\times 10^2$ & $-59$ & $30$ & $-75$ & $-84$\\

Poland & POL & $6.01(2)$ & $0.588(15)$ & $9.6(4)\times 10^2$ & $-41$ & $17$ & $-59$ & $-70$\\

Ireland & IRL & $8.541(7)$ & $0.612(11)$ & $9.8(5)\times 10^2$ & $-60$ & $25$ & $-61$ & $-66$\\

Netherlands & NLD & $7.884(6)$ & $0.555(7)$ & $277(5)$ & $-43$ & $16$ & $-48$ & $-47$\\

Romania & ROU & $6.877(17)$ & $0.461(12)$ & $176(8)$ & $-46$ & $17$ & $-66$ & $-73$\\

Bulgaria & BGR & $6.13(2)$ & $0.326(8)$ & $40(2)$ & $-42$ & $16$ & $-64$ & $-68$\\

Croatia & HRV & $6.25(1)$ & $0.78(2)$ & $4.7(2)\times 10^3$ & $-52$ & $19$ & $-66$ & $-64$\\

Norway & NOR & $7.261(6)$ & $0.710(13)$ & $0.84(2) \times 10^{3}$ & $-46$ & $16$ & $-30$ & $-37$\\

Austria & AUT & $7.410(6)$ & $1.07(2)$ & $5.9(2)\times 10^4$ & $-53$ & $20$ & $-56$ & $-64$\\

Hungary & HUN & $5.961(11)$ & $0.55(1)$ & $6.5(3)\times 10^2$ & $-42$ & $17$ & $-48$ & $-69$\\

Serbia & SRB & $7.399(7)$ & $0.658(9)$ & $2.68(9)\times 10^3$ & $-59$ & $20$ & $-68$ & $-71$\\

Belgium & BEL & $8.498(8)$ & $0.585(8)$ & $338(7)$ & $-57$ & $24$ & $-56$ & $-43$\\
 \hline
 \end{tabular}}
 \caption{{\bf First wave fits and average mobility reductions in Europe}. Values of the fit for $a$, $b$ and $\gamma$ for the first wave in the 21 European countries considered in this study, together with the 95\% CL error. Six week average mobility reduction for Google and Apple categories.}
 \label{T1}
 \end{center}
 \end{table*}
 
 \begin{table*}[tb!]
\begin{center}
{\footnotesize \begin{tabular}{ || l |c|c|c|c|c|c||}
\hline
\multicolumn{7}{|c|}{Fit parameters: two-gamma function parameters in Europe}\\
\hline
Country & ISO & $a$ & $\gamma_B$ & $\gamma_C$ & $b$ & $\Delta$t \\
\hline
France & FRA & $7.663(9)$ & $0.715(13)$ & $0.64(2)$ & $1.65(4)\times 10^3$ & $2.6(2)$\\

Italy & ITA & $8.304(9)$ & $0.591(6)$ & $0.386(8)$ & $215(3)$ & $2.88(4)$\\

Spain & ESP & $8.483(9)$ & $0.79(1)$ & $0.62(2)$ & $2.68(5)\times 10^3$ & $2.74(8)$\\

United Kingdom & GBR & $8.339(4)$ & $0.549(9)$ & $0.363(3)$ & $429(9)$ & $2.78(5)$\\

Germany & DEU & $7.637(15)$ & $0.726(13)$ & $0.57(3)$ & $1.75(4)\times 10^3$ & $2.84(11)$\\

Denmark & DNK & $8.06(11)$ & $0.38(2)$ & $0.18(2)$ & $24(1)$ & $4.60(6)$\\

Switzerland & CHE & $8.158(7)$ & $0.80(1)$ & $0.71(3)$ & $2.53(4)\times 10^3$ & $2.8(2)$\\

Sweden & SWE & $8.51(11)$ & $0.353(12)$ & $0.26(2)$ & $39(1)$ & $3.57(8)$\\

Slovakia & SVK & $6.04(7)$ & $0.56(25)$ & $0.43(3)$ & $5(2)\times 10^2$ & $1.4(1.2)$\\

Portugal & PRT & $7.95(1)$ & $0.80(2)$ & $0.548(14)$ & $6.7(4)\times 10^3$ & $2.61(8)$\\

Poland & POL & $6.25(3)$ & $0.622(12)$ & $0.416(16)$ & $1.70(5)\times 10^3$ & $3.07(5)$\\

Ireland & IRL & $8.70(4)$ & $0.52(2)$ & $0.26(4)$ & $337(15)$ & $6.26(9)$\\

Netherlands & NLD & $7.900(5)$ & $0.67(3)$ & $0.528(7)$ & $1.18(5)\times 10^3$ & $1.91(14)$\\

Romania & ROU & $7.04(2)$ & $0.557(13)$ & $0.34(1)$ & $7.0(3)\times 10^2$ & $3.87(7)$\\

Bulgaria & BGR & $6.15(2)$ & $0.86(31)$ & $0.32(8)$ & $1.9(7)\times 10^4$ & $1.3(2)$\\

Croatia & HRV & $6.296(8)$ & $1.04(4)$ & $0.672(15)$ & $1.07(7)\times 10^5$ & $1.89(7)$\\

Norway & NOR & $7.42(4)$ & $0.62(2)$ & $0.32(4)$ & $349(6)$ & $3.54(5)$\\

Austria & AUT & $7.462(12)$ & $1.060(17)$ & $0.75(5)$  & $6.1(2)\times 10^4$ & $2.35(7)$\\

Hungary & HUN & $5.972(11)$ & $0.95(22)$ & $0.54(1)$ & $8(3)\times 10^4$ & $1.5(2)$\\

Serbia & SRB & $7.44(3)$ & $0.641(14)$ & $0.55(5)$ & $2.18(8)\times 10^3$ & $5.14(18)$\\

Belgium & BEL & $8.481(17)$ & $0.585(11)$ & $0.62(4)$ & $335.15(8)$ & $3.7(6)$\\
 \hline
 \end{tabular}}
 \caption{{\bf First wave fits for the two-gamma model for Europe}. Outcome of the two-gamma fits for the 5 parameters $a$, $\gamma_B$, $\gamma_C$, $b$ and $\Delta t$. The errors refer to a 95\% CL.}
 \label{T2}
 \end{center}
 \end{table*}

 \begin{table*}[tb!]
\begin{center}
{\footnotesize \begin{tabular}{ || l |c|c|c|c|c|c|c||}
\hline
\multicolumn{8}{|c|}{Fit parameters: first wave and 6-week average mobility variations in the US}\\
\hline
State &  & $a$ & $\gamma$ & $b$ & Work. & Res. & Driv.\\
\hline
Alabama & AL & $7.92(5)$ & $0.40(2)$ & $74(4)$ & $-35$ & $14$ & $-23$\\

Alaska & AK & $6.196(7)$ & $0.89(2)$ & $2.1(2)\times 10^4$ & $-36$ & $14$ & $-23$\\

Arizona & AZ & $8.28(7)$ & $0.266(12)$ & $17.5(6)$ & $-40$ & $15$ & $-36$\\

Arkansas & AR & $7.72(9)$ & $0.37(2)$ & $63(3)$ & $-32$ & $11$ & $-20$\\

California & CA & $8.44(4)$ & $0.243(8)$ & $12.3(5)$ & $-46$ & $20$ & $-48$\\

Colorado & CO & $8.01(1)$  & $0.331(6)$ & $31(1)$ & $-47$ & $18$ & $-46$\\

Connecticut & CT & $9.124(4)$ & $0.413(6)$ & $73(3)$ & $-45$ & $19$ & $-40$\\

Delaware & DE & $7.577(7)$ & $0.399(5)$ & $121(4)$ & $-43$ & $17$ & $-35$\\

Florida & FL & $9.76(2)$ & $0.335(14)$ & $20(2)$ & $-42$ & $17$ & $-46$\\

Georgia & GA & $11.21(2)$ & $0.316(9)$ & $16.3(7)$ & $-41$ & $16$ & $-32$\\

Hawaii & HI & $6.578(2)$ & $0.919(11)$ & $2.81(8)\times 10^4$ & $-45$ & $20$ & $-62$\\

Idaho & ID & $7.751(16)$ & $0.71(4)$ & $2.2(3)\times 10^3$ & $-38$ & $12$ & $-15$\\

Illinois & IL & $8.93(3)$ & $0.322(8)$ & $34(2)$ & $-45$ & $19$ & $-38$\\

Indiana & IN & $8.312(12)$ & $0.331(6)$ & $35(2)$ & $-41$ & $16$ & $-30$\\

Iowa & IA & $9.721(13)$ & $0.415(7)$ & $169(8)$ & $-34$ & $14$ & $-22$\\

Kansas & KS & $9.12(7)$ & $0.335(17)$ & $53(4)$ & $-37$ & $14$ & $-29$\\

Kentucky & KY & $6.71(3)$ & $0.395(13)$ & $120(6)$ & $-39$ & $14$ & $-26$\\

Louisiana & LA & $8.424(11)$ & $0.82(3)$ & $8(1)\times 10^3$ & $-39$ & $15$ & $-46$\\

Maine & ME & $7.237(11)$ & $0.195(3)$ & $6.7(2)$ & $-39$ & $15$ & $-35$\\

Maryland & MD & $10.170(7)$ & $0.293(3)$ & $21.2(5)$ & $-47$ & $20$ & $-40$\\

Massachusetts & MA & $10.110(3)$ & $0.405(4)$ & $64(2)$ & $-51$ & $22$ & $-48$\\

Michigan & MI & $10.820(9)$ & $0.359(8)$ & $14.6(6)$ & $-51$ & $20$ & $-46$\\

Minnesota & MN & $8.727(11)$ & $0.376(6)$ & $137(6)$ & $-43$ & $19$ & $-41$\\

Mississippi & MS & $7.88(2)$ & $0.324(7)$ & $36(1)$ & $-34$ & $13$ & $-20$\\

Missouri & MO & $7.11(2)$ & $0.447(18)$ & $132(8)$ & $-38$ & $14$ & $-32$\\

Montana & MT & $4.390(5)$ & $1.18(3)$ & $1.1(7)\times 10^5$ & $-36$ & $13$ & $-15$\\

Nebraska & NE & $8.75(8)$ & $0.409(7)$ & $233(12)$ & $-33$ & $14$ & $-20$\\

Nevada & NV & $7.08(2)$ & $0.467(16)$ & $149(9)$ & $-50$ & $18$ & $-38$\\

New Hampshire & NH & $8.704(4)$ & $0.285(2)$ & $18.8(3)$ & $-42$ & $18$ & $-37$\\

New Jersey & NJ & $11.340(5)$ & $0.493(7)$ & $127(4)$ & $-52$ & $23$ & $-47$\\

New Mexico & NM & $8.06(2)$ & $0.346(7)$ & $59(2)$ & $-39$ & $15$ & $-33$\\

New York & NY & $12.530(3)$ & $0.514(6)$ & $110(4)$ & $-53$ & $22$ & $-51$\\

North Carolina & NC & $10.73(3)$ & $0.1100(16)$ & $3.55(3)$ & $-38$ & $14$ & $-28$\\

North Dakota & ND & $5.22(1)$ & $0.358(6)$ & $151(6)$ & $-32$ & $16$ & $-25$\\

Ohio & OH & $9.993(14)$ & $0.310(7)$ & $21.4(7)$ & $-42$ & $16$ & $-30$\\

Oklahoma & OK & $6.45(3)$ & $0.345(14)$ & $45(3)$ & $-36$ & $13$ & $-27$\\

Oregon & OR & $6.68(2)$ & $0.53(2)$ & $2.8(3)\times 10^2$ & $-42$ & $15$ & $-37$\\

Pennsylvania & PA & $9.8(1)$ & $0.387(9)$ & $45(3)$ & $-46$ & $19$ & $-43$\\

Rhode Island & RI & $7.155(5)$  & $0.464(6)$ & $2.7(1)\times 10^2$ & $-43$ & $18$ & $-38$\\

South Carolina & SC & $9.21(3)$ & $0.376(15)$ & $39(3)$ & $-36$ & $13$ & $-25$\\

South Dakota & SD & $6.041(12)$ & $0.93(2)$ & $2.1(2)\times 10^5$ & $-31$ & $14$ & $-13$\\

Tennessee & TN & $6.99(3)$ & $0.290(9)$ & $25(1)$ & $-38$ & $14$ & $-30$\\

Texas & TX & $9.79(4)$ & $0.379(16)$ & $48(3)$ & $-41$ & $17$ & $-38$\\

Utah & UT & $9.69(4)$ & $0.320(12)$ & $20.4(7)$ & $-41$ & $16$ & $-23$\\

Vermont & VT & $4.662(5)$ & $1.03(2)$ & $1.6(2)\times 10^5$ & $-47$ & $18$ & $-48$\\

Virginia & VA & $10.64(1)$ & $0.264(3)$ & $16.0(4)$ & $-42$ & $17$ & $-35$\\

Washington & WA & $7.896(13)$ & $0.382(9)$ & $18(2)$ & $-48$ & $18$ & $-32$\\

West Virginia & WV & $8.04(4)$ & $0.33(2)$ & $26(2)$ & $-37$ & $13$ & $-28$\\

Wisconsin & WI & $8.93(4)$ & $0.204(6)$ & $7.7(2)$ & $-39$ & $17$ & $-31$\\

Wyoming & WY & $7.60(4)$ & $0.296(15)$ & $20(1)$ & $-33$ & $13$ & $-10$\\
 \hline
 \end{tabular}}
 \caption{{\bf First wave fits and average mobility reductions in the US}. Same as Table~\ref{T1}.}
 \label{T3}
 \end{center}
 \end{table*}
 
 \begin{table*}[tb!]
\begin{center}
{\footnotesize \begin{tabular}{ || l |c|c|c|c|c|c||}
\hline
\multicolumn{7}{|c|}{Fit parameters: two-gamma function parameters in the US}\\
\hline
State &  & $a$ & $\gamma_B$ & $\gamma_C$ & $b$ & $\Delta$t \\
\hline
Alaska & AK  & $ 6.219(6) $ & $ 1.28(8) $ & $ 0.80(2) $ & $ 2.9(3)\times 10^6 $ & $ 1.52(8)$\\

Arizona & AZ & $8.87(7)$ & $0.48(3)$ & $0.186(6)$ & $3.7(3)\times 10^3$ & $2.87(6)$\\

Arkansas & AR & $7.9(2)$ & $0.9(4)$ & $0.33(3)$ & $2.2(8)\times 10^4$ & $1.2(3)$\\

California & CA & $8.80(5)$ & $0.41(3)$ & $0.179(6)$ & $1.5(2)\times 10^2$ & $3.6(1)$\\

Colorado & CO & $8.056(9)$ & $0.63(6)$ & $0.301(5)$ & $1.7(2)\times 10^3$ & $2.7(2)$\\

Connecticut & CT & $9.156(4)$ & $0.480(8)$ & $0.333(7)$ & $215(8)$ & $5.3(2)$\\

Delaware & DE & $7.596(7)$ & $0.59(5)$ & $0.379(6)$ & $2.1(3)\times 10^3$ & $3.6(2)$\\

Florida & FL & $10.16(5)$ & $0.51(3)$ & $0.172(9)$ & $2.6(2)\times 10^2$ & $3.55(7)$\\

Georgia & GA & $11.40(2)$ & $0.45(2)$ & $0.228(6)$ & $126(6)$ & $3.86(8)$\\

Hawaii & HI & $6.586(7)$ & $0.91(2)$ & $0.7(2)$ & $2.89(9)\times 10^4$ & $4.9(4)$\\

Idaho & ID & $7.95(2)$ & $1.23(5)$ & $0.31(2)$ & $2.6(3)\times 10^6$ & $2.66(3)$\\

Illinois & IL & $9.07(3)$ & $0.53(4)$ & $0.283(6)$ & $6.3(6)\times 10^2$ & $2.8(1)$\\

Indiana & IN & $8.378(8)$ & $0.62(3)$ & $0.293(4)$ & $1.9(3)\times 10^3$ & $3.01(6)$\\

Iowa & IA & $10.3(1)$ & $0.33(2)$ & $0.18(2)$ & $62(2)$ & $7.38(5)$\\

Kansas & KS & $9.14(9)$ & $0.6(6)$ & $0.33(2)$ & $1(1)\times 10^3$ & $1.9(1.3)$\\

Kentucky & KY & $6.66(9)$ & $0.40(3)$ & $0.5(1)$ & $130(6)$ & $6.9(6)$\\

Louisiana & LA & $8.65(3)$ & $0.91(3)$ & $0.31(3)$ & $3.7(2)\times 10^4$ & $3.35(4)$\\

Maine & ME & $7.279(9)$ & $0.53(6)$ & $0.180(3)$ & $5.7(7)\times 10^3$ & $2.6(2)$\\

Maryland & MD & $10.197(6)$ & $0.48(3)$ & $0.277(3)$ & $3.3(3)\times 10^2$ & $3.7(2)$\\

Massachusetts & MA & $10.138(4)$ & $0.441(6)$ & $0.339(6)$ & $120(3)$ & $5.8(2)$\\

Michigan & MI & $10.902(6)$ & $0.47(2)$ & $0.274(4)$ & $66(2)$ & $2.55(7)$\\

Minnesota & MN & $8.695(8)$ & $0.23(2)$ & $0.400(5)$ & $16(1)$ & $4.9(2)$\\

Mississippi & MS & $8.00(2)$ & $0.50(4)$ & $0.284(6)$ & $4.0(3)\times 10^2$ & $2.9(2)$\\

Missouri & MO & $7.32(3)$ & $0.66(3) $ & $ 0.30(2) $ & $ 2.9(2)\times 10^3 $ & $ 3.07(7)$\\

Montana & MT & $4.389(6)$ & $1.18(3) $ & $ 0.767(0) $ & $ 1.33(9)\times 10^6 $ & $ 7.7(0)$\\

Nebraska & NE & $8.748(9)$ & $0.36(9) $ & $ 0.411(8) $ & $ 1.2(4)\times 10^2 $ & $ 4(2)$\\

Nevada & NV & $7.24(2)$ & $0.73(3) $ & $ 0.333(9) $ & $ 5.2(4)\times 10^3 $ & $ 2.89(5)$\\

New Hampshire & NH & $8.705(4)$ & $0.284(3) $ & $ 0.095(0) $ & $ 19.5(3) $ & $ 19(0)$\\

New Jersey & NJ & $11.361(3)$ & $0.71(2) $ & $ 0.447(4) $ & $ 2.3(1)\times 10^3 $ & $ 2.83(6)$\\

New Mexico & NM & $8.18(3)$ & $0.49(3) $ & $ 0.306(6) $ & $ 4.8(4)\times 10^2 $ & $ 3.6(1)$\\

New York & NY & $12.529(4)$ & $0.513(7) $ & $ 0.0855(0) $ & $ 115(3) $ & $ 18(0)$\\

North Carolina & NC & $10.88(3)$ & $0.28(3) $ & $ 0.103(1) $ & $ 43(4) $ & $ 3.9(2)$\\

North Dakota & ND & $5.26(2)$ & $0.45(3) $ & $ 0.331(8) $ & $ 7.0(8)\times10^3 $ & $ 4.9(3)$\\

Ohio & OH & $10.18(3)$ & $0.351(7) $ & $ 0.222(8) $ & $ 46(2) $ & $ 5.29(9)$\\

Oklahoma & OK & $6.71(3)$ & $0.68(3) $ & $ 0.230(7) $ & $ 4.7(4)\times 10^3 $ & $ 3.17(5)$\\

Oregon & OR & $7.02(5)$ & $0.70(3) $ & $ 0.30(2) $ & $ 2.0(2)\times 10^3 $ & $ 2.80(6)$\\

Pennsylvania & PA & $9.942(5)$ & $0.60(2) $ & $ 0.296(4) $ & $ 9.5(4)\times 10^2 $ & $ 3.88(4)$\\

Rhode Island & RI & $7.195(4)$ & $0.533(7) $ & $ 0.384(6) $ & $ 8.4(3)\times 10^2 $ & $ 5.47(8)$\\

South Carolina & SC & $9.57(5)$ & $0.50(2) $ & $ 0.23(2) $ & $ 2.7(2)\times 10^2 $ & $ 3.17(7)$\\

South Dakota & SD & $6.04(2)$ & $0.93(3) $ & $ 0.333(0) $ & $ 2.4(2)\times 10^5 $ & $ 8.7(0)$\\

Tennessee & TN & $7.3(2)$ & $0.25(2) $ & $ 0.18(4) $ & $ 18.0(7) $ & $ 6.6(3)$\\

Texas & TX & $10.56(9)$ & $0.43(2) $ & $ 0.19(1) $ & $ 147(5) $ & $ 3.79(4)$\\

Utah & UT & $10.01(4)$ & $0.48(6) $ & $ 0.232(6) $ & $ 208(8) $ & $ 2.89(5)$\\

Vermont & VT & $4.75(2)$ & $0.98(3) $ & $ 0.36(6) $ & $ 1.06(5)\times 10^6 $ & $ 3.72(6)$\\

Virginia & VA & $10.67(1)$ & $0.43(6) $ & $ 0.256(4) $ & $ 1.7(3)\times 10^2 $ & $ 3.4(3)$\\

Washington & WA & $8.20(2)$ & $0.349(5) $ & $ 0.204(5) $ & $ 15.3(2) $ & $ 2.36(4)$\\

West Virginia & WV & $9.4(5)$ & $0.35(4) $ & $ 0.09(2) $ & $ 67(5) $ & $ 4.15(8)$\\

Wisconsin & WI & $9.13(3)$ & $0.45(4) $ & $ 0.176(4) $ & $ 2.2(2)\times 10^2 $ & $ 2.6(1)$\\

Wyoming & WY & $7.9(3)$ & $0.25(4) $ & $ 0.17(7) $ & $ 13.0(7) $ & $ 7.3(5)$\\
 \hline
 \end{tabular}}
 \caption{{\bf First wave fits for the two-gamma model for the US}. Same as Table~\ref{T2}.}
 \label{T4}
 \end{center}
 \end{table*}

\end{document}